\documentclass[preprint,12pt]{elsarticle}

\usepackage{amssymb}

\usepackage{enumitem}
\usepackage{multirow}
\usepackage{booktabs}
\usepackage{array}
\usepackage{makecell}
\usepackage{graphicx}
\usepackage{xcolor}
\usepackage{hyperref}
\usepackage{comment}
\usepackage{rotating}
\usepackage[version=3]{mhchem}
\biboptions{sort&compress}
\usepackage{tabularx}
\journal{}
\usepackage{ragged2e}
\usepackage{multicol}
\usepackage{makecell}
\usepackage{blindtext}
\usepackage[left=2cm,right=2cm]{geometry}
\usepackage{caption}
\newcommand{\fullsizefigure}[2]{%
  \begin{figure*}[!htbp] 
    \centering
    \includegraphics[keepaspectratio]{#1}
    \caption{#2}
    \label{#1}
  \end{figure*}
}

\newcommand{\halfsizefigure}[2]{%
  \begin{figure}[!htbp]
    \centering
    \includegraphics[keepaspectratio]{#1}
    \caption{#2}
    \label{#1}
  \end{figure}
}

\begin{document}

\begin{frontmatter}



\title{A Computational Study for Screening High-Selectivity Inhibitors in Area-Selective Atomic Layer Deposition on Amorphous Surfaces}


\author[inst1]{Gijin Kim\fnref{equal}}
\author[inst1]{Purun-hanul Kim\fnref{equal}}
\author[inst2]{Suk Gyu Hahm}
\author[inst2]{Myongjong Kwon}
\author[inst2]{Byungha Park}
\author[inst1]{Changho Hong\corref{cor2}}
\author[inst1,inst3,inst4]{Seungwu Han\corref{cor2}}

\fntext[equal]{These authors contributed equally to this work}
\cortext[cor2]{Corresponding authors}

\affiliation[inst1]{organization={Department of Materials Science and Engineering},
            addressline={Seoul National University}, 
            city={Seoul},
            postcode={08826}, 
            country={Republic of Korea}}
\affiliation[inst2]{organization={Samsung Advanced Institute of Technology},
            addressline={Samsung Electronics}, 
            city={Suwon},
            postcode={16678}, 
            country={Republic of Korea}}
\affiliation[inst3]{addressline={AI center, Korea Institute for Advanced Study}, 
            city={Seoul},
            postcode={02455}, 
            country={Republic of Korea}}
\affiliation[inst4]{addressline={Research Institue of Advanced Materials, Seoul National University}, 
            city={Seoul},
            postcode={08826}, 
            country={Republic of Korea}}

\begin{abstract}

Area-selective atomic layer deposition (AS-ALD) is an emerging technology in semiconductor manufacturing. However, accurately understanding inhibitor reactivity on surfaces remains challenging particularly when the substrate is amorphous. In this study, we employ density functional theory (DFT) to investigate reaction pathways and quantify the reactivity of (N,N-dimethylamino)trimethylsilane (DMATMS) and ethyltrichlorosilane (ETS) at silanol (\ce{-OH}), siloxane (\ce{-O-}), amine (\ce{-NH2}), and imide (\ce{-NH-}) sites on both amorphous and crystalline silicon oxide and silicon nitride surfaces. Notably, both molecules exhibit greater reactivity toward terminal sites (\ce{-OH} and \ce{-NH2}) on amorphous surfaces compared to crystalline counterparts. For bridge sites, \ce{-O-} and \ce{-NH-}, multiple reaction pathways are identified, with bridge-cleavage reactions being the predominant mechanism, except for DMATMS reactions with nitride surfaces. The reactivity of DMATMS with \ce{-NH-} sites is comparable to that with \ce{-NH2}, with both reactions yielding volatile products. This study underscores the importance of amorphous surface modeling in reliably predicting inhibitor adsorption and reactivity on realistic surfaces. Moreover, we outline a computational screening approach that accounts for site-specific precursor–inhibitor interactions, enabling efficient and rational theoretical design of AS-ALD precursor–inhibitor pairs.

\end{abstract}

\begin{graphicalabstract}
\includegraphics{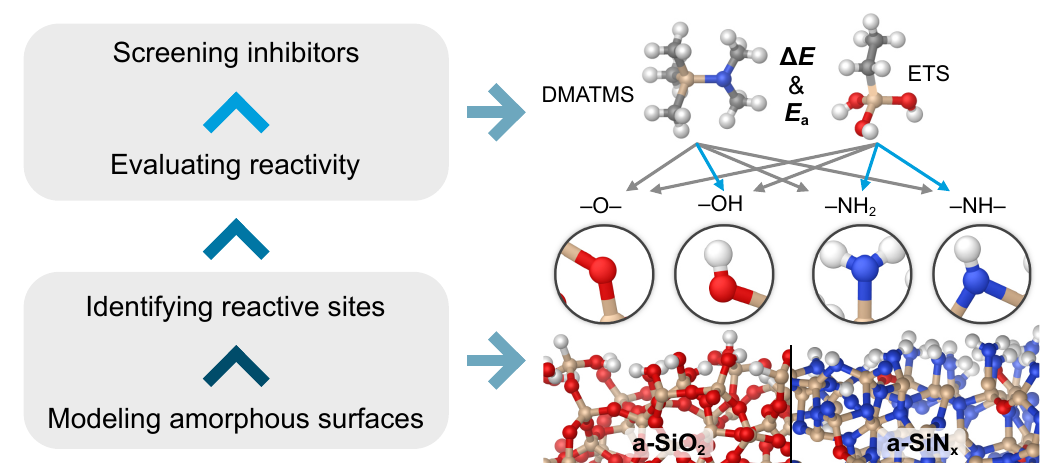}
\end{graphicalabstract}

\begin{highlights}
\item Amorphous surface models provide more realistic and accurate reactivity predictions.
\item The amorphous nature of surfaces enhances inhibitor reactivity.
\item A DFT-based screening protocol is proposed for the rational design of AS-ALD systems.
\end{highlights}


\end{frontmatter}


\section{Introduction}
\label{sec:Introduction}


    The semiconductor industry has continually pursued enhanced device performance and reduced manufacturing costs through shrinking feature sizes~\cite{Clark.2018}. As device scaling moves beyond the 10~nm node, conventional patterning techniques such as photolithography combined with atomic layer deposition (ALD), frequently necessitate multipatterning. This approach involves repeated cycles of lithography and etching to overcome resolution limitations inherent to photolithography, even when employing advanced methods such as extreme ultraviolet (EUV)~\cite{Raley.2019,Levinson.2022}. However, multipatterning  elevates process complexity, manufacturing costs, and alignment precision requirements, thus increasing the likelihood of edge-placement errors (EPE) and subsequent yield loss~\cite{Zhang.2022,Yasmeen.2022}. These challenges are particularly critical in advanced device architectures, including high-density 3D NAND, FinFET, and Gate-All-Around structures, which demand uniform thin-film deposition and defect-free patterning on nanoscale three-dimensional surfaces.

    To overcome the challenges in the above, bottom-up approaches such as area-selective atomic layer deposition (AS-ALD) have attracted growing interest~\cite{Parsons.2020,Mackus.2019,Zhang.2022,Yasmeen.2022}. AS-ALD exploits the chemical selectivity of precursors between different surfaces, with the selectivity facilitated by deposited inhibitors, thereby enabling material deposition exclusively onto the targeted growth surface (GS) while preserving the adjacent non-growth surface (NGS). By integrating deposition and patterning into a single process, AS-ALD significantly reduces the complexity associated with multipatterning techniques. Furthermore, AS-ALD minimizes EPE by directly depositing materials onto GSs. These advantages make AS-ALD particularly promising for fabricating semiconductor features below 10 nm. 

    While AS-ALD typically relies on inhibitors to enhance selectivity, certain systems, such as \ce{SiO2} deposition~\cite{Lee.2021s4,Lee.2022}, can accomplish selective deposition using the intrinsic selective reactivity of precursors in the absence of inhibitor molecules~\cite{Cao.2020}. These inherent AS-ALD approaches offer the advantage of streamlined processing by eliminating the need for additional species introduction. However, such systems often necessitate additional surface modifications and tuning of process parameters including temperature and partial pressure. These stringent requirements can result in compromised performance, such as reduced growth rates or deteriorated film quality, limiting the practical implementation of intrinsic selectivity to only a few demonstrated cases.~\cite{Cao.2020,Mameli.2023,Yarbrough.2023ka3}.

    Inhibitor-based AS-ALD involves selective inhibitor deposition onto the NGS prior to precursor deposition. In early process developments, self-assembled monolayers (SAMs) were adopted as inhibitors. SAMs possess long carbon backbone chains that lead to dense molecular packing and strong van der Waals interactions. Octadecylphosphonic acid (ODPA)~\cite{Bobb-Semple.2019}, undecylaldehyde (UDA)~\cite{Park.20248xr}, and octadecyltrichlorosilane (ODTS)~\cite{Liu.2021} have been investigated in AS-ALD systems. However, ongoing device miniaturization and the slow inhibitor deposition rates of SAMs have driven the introduction of small-molecule inhibitors (SMIs) such as (N,N-dimethylamino)trimethylsilane (DMATMS)~\cite{Dongen.2023,Soethoudt.2020,Soethoudt.2019,Khan.2018} and acetylacetone (hacac)~\cite{Mameli.2017,Merkx.2020,Merkx.2022}, among others~\cite{Kim.2020,Yarbrough.2023ka3,Gasvoda.2021,Karasulu.2023}. Unlike SAMs, SMIs are characterized by smaller molecular structures without extended backbone chains, offering improved compatibility with shrinking feature sizes and faster deposition rates. 

    The performance of these SMIs is strongly influenced by their head groups, which anchor the NGS~\cite{Yarbrough.2021}. Specifically, head groups affect the selective passivation of NGSs, passivation stability across relevant temperature windows, and processing times required for achieving complete surface coverage. These performance characteristics originate from the chemical specificity and surface-binding properties defined by interactions between inhibitor head groups and surface functional groups~\cite{Yu.2024ma,Merkx.2022}. Currently, prediction of inhibitor–surface interactions often rely on empirical evaluations based on acid/base characteristics; however, such simplified approaches rarely provide universally predictive or accurate guidelines for designing AS-ALD systems~\cite{Karasulu.2023, Bobb-Semple.2019}. Consequently, inhibitor development continues to depend heavily on labor-intensive trial-and-error methods, highlighting a substantial gap in the fundamental understanding required for predictive inhibitor design.
    
    These challenges are especially pronounced in systems requiring selective treatment of chemically similar GS and NGS, such as SiO$_2$ and Si$_3$N$_4$. The selective deposition on these materials is critical for 3D NAND flash memory structures, where conventional patterning approaches face limitations in achieving the required precision and aspect ratios, driving the need for AS-ALD-based solutions~\cite{Arts.2022}. Most inhibitors used in these systems have been adapted from or inspired by existing ALD precursors. For example, aminosilanes, chlorosilanes, and alkoxysilanes are employed on SiO$_2$ surfaces, whereas aldehydes and cyclic azasilanes are employed on Si$_3$N$_4$ surfaces~\cite{Gasvoda.2021, Park.20248xr}. However, due to minimal differences in reaction mechanisms between the inhibitor head groups and their respective surface chemistries, achieving higher selectivity remains challenging. Although selective deposition is attainable at lower temperatures, it generally leads to prolonged processing times~\cite{Ovanesyan.2019, Murray.2014}. Currently, developing inhibitors largely rely on empirical methods and is limited to precursor chemistries previously established in ALD processes, entailing substantial costs for equipment and precursor synthesis~\cite{Kim.2024hgs}.

    In response, computational methods, particularly density functional theory (DFT), have emerged as useful tools providing mechanistic insights into inhibitor-surface interactions. DFT has been extensively applied to investigate surface reactions in ALD~\cite{Kang.2021326,Sandupatla.2015,Lee.2020,Jeong.2013,Kim.2014,Huang.2017,Yang.2014,Roh.2022} and AS-ALD~\cite{Tezsevin.2023,Yu.2024ma,Yarbrough.2022,Lee.2021s4,Lee.2022}, through analysis of reaction energies and activation energies along adsorption pathways.

    Existing computational studies, however, typically utilize idealized crystalline surface models, which can lead to discrepancies with experimental observations on amorphous substrates. For example, previous DFT calculations reported relatively high reaction barriers of $\sim$1 eV for aminosilanes on perfect crystalline silicon nitride, suggesting negligible reactivity at 150 $^\circ$C~\cite{Murray.2014, Huang.2014}. Nevertheless, experiments have demonstrated adsorption of aminosilanes onto amorphous \ce{SiN}$_x$ surfaces at 150 $^\circ$C~\cite{Xu.2022}. Similarly, DFT calculations indicated kinetically unfavorable barriers (1.78 eV) for pivalic acid adsorption on crystalline SiO$_2$, while experiments showed adsorption occurring readily~\cite{Karasulu.2023}. This discrepancy is largely attributed to structural differences between crystalline and amorphous substrate models. Indeed, experimental amorphous substrates such as SiO$_2$ and Si$_3$N$_4$ exhibit disordered reactive site distributions with reactive bridge sites, lower atomic densities, and varied hydrogen-bonding environments, significantly impacting local adsorption energies and reaction barriers~\cite{Yarbrough.2022, Soethoudt.2019, Merkx.2022}. Accordingly, detailed analysis of diverse surface functional groups and their corresponding reaction pathways with inhibitors is critical for achieving an accurate understanding of inhibitor–surface interactions and for formulating reliable computational models for AS-ALD processes.

    In this work, we propose a computational approach for evaluating inhibitor–surface reactivity in AS-ALD systems, accounting for structural differences between crystalline and amorphous surfaces. We quantify inhibitor–surface interactions by evaluating reaction and activation energies associated with adsorption pathways involving functional groups characteristic of diverse local motifs found in amorphous surfaces.
    We focus on silicon oxide (\ce{SiO2}) and silicon nitride (\ce{Si3N4}) surfaces, examining two representative small-molecule inhibitors (SMIs): DMATMS and ethyltrichlorosilane (ETS). DMATMS is an aminosilane containing a reactive amine head group and an inert alkyl tail group. In contrast, ETS possesses a reactive Si–Cl head group analogous to that of octadecyltrichlorosilane (ODTS)~\cite{Wang.2003,Saner.2011,Roscioni.2016,Xu.2004}, a widely used SAM inhibitor, but with a significantly shorter alkyl chain. This selection enables us to isolate the effect of the head group independently from alkyl chain length and molecular packing effects.

\section{Methods}
\label{sec:Methods}

All DFT calculations in this work are conducted with the Vienna \textit{Ab initio} Simulation Package (VASP), which utilizes the projector augmented wave (PAW) scheme. We adopt the Perdew-Burke-Ernzerhof (PBE) functional to describe the exchange-correlation energy~\cite{Kresse.1996,Kresse.19962o,Perdew.1996, Perdew.1996ifi}.
For self-consistent field (SCF) calculations, we use the following computational settings: a maximum plane-wave kinetic energy cutoff of 480 eV, $\Gamma$-only \textit{k}-point sampling, Gaussian smearing of 0.05 eV width, and an energy convergence criterion of $10^{-4}$ eV. Structural relaxation is performed via the conjugate gradient (CG) algorithm until forces on atoms are below 0.03 eV/\AA.
Since van der Waals (vdW) interactions are crucial for  physisorption, we employ Grimme's DFT-D3 dispersion-correction scheme with the Becke-Johnson (BJ) damping function~\cite{Grimme.2010, Grimme.2011}. For slab calculations, dipole corrections are applied along the z-direction to compensate for the dipole moment induced by asymmetry of the amorphous slab.

Activation energies are determined through the climbing-image nudged elastic band (CI-NEB) method as implemented in the VASP Transition State Tools (VTST) code~\cite{Henkelman.2000,Henkelman.2000dvp}. Seven images, including the initial and final configurations, are involved for CI-NEB calculations, with the initial and final configurations pre-relaxed. Structures of intermediate images are optimized by the fast inertial relaxation engine (FIRE)~\cite{Bitzek.2006} and limited-memory Broyden-Fletcher-Goldfarb-Shanno (LBFGS)~\cite{Nocedal.1980} optimizers with a force convergence criterion of 0.05 eV/\AA. For reactions with multiple barriers, we subdivide the pathway to identify individual barriers, taking the highest activation energy as the rate-determining step.

All \textit{ab initio} molecular dynamics (AIMD) simulations are performed in the NVT ensemble with the Nos\'{e}--Hoover thermostat~\cite{Hoover.1985}. For melt-quench simulations to generate the amorphous slab, a timestep of 2 fs is used for all processes except for the pre-melting stage, conducted with a timestep of 1 fs. To reduce computational cost while simulating hydrogen on the surface, we employ increased hydrogen masses in the MD simulations without using smaller timesteps. For temperatures above 500 K, we use a mass of 12 amu, and for temperatures below 500 K, we use 3 amu. All MD simulations utilize a cutoff energy of 320 eV and soft pseudopotentials with larger core radii to reduce computational expense. Following the MD simulations, the final structures are re-optimized with the structural relaxation parameters described above (480 eV plane-wave cutoff, $\Gamma$-only \textit{k}-point sampling, and 0.03 eV/\AA\ force convergence criterion) to prepare them for subsequent calculations.

Since chlorosilanes are known to readily undergo hydrolysis upon exposure to moisture~\cite{Eakins.1968,Hair.1995}, our computational results indicate that ETS reacts with residual humidity typically present in deposition chambers (see supplementary information S1) Consequently, we model the hydrolyzed form of ETS rather than its pristine molecular structure. Hereafter, ETS refers to the hydrolyzed form unless otherwise noted.

\section{Results}
\label{sec:Results and discussion}

\subsection{Surface modeling}

In this section, we atomistically model amorphous slabs that exhibit experimentally observed properties. We model plasma-enhanced chemical vapor deposition (PECVD)-grown amorphous silicon oxide (a-SiO$_2$) and silicon nitride (a-SiN$_x$) substrates, utilizing bulk properties representative of common semiconductor-industry PECVD recipes. For a-SiO$_2$, we adopt a bulk density of 2.2 g/cm$^3$ and a bulk Si:O atomic ratio of 1:2~\cite{Deenapanray.1998,Pan.1985,Ray.1996,Bulla.1998}, and for a-SiN$_x$, a bulk density of 2.6 g/cm$^3$ with a bulk Si:N atomic ratio of 9:11~\cite{Lin.1998,Park.2001,Kaloyeros.2020sg}. PECVD-grown a-SiN$_x$ films can be particularly susceptible to oxygen incorporation under fab environments, leading to silicon oxynitride (SiON) formation~\cite{García.1995,Boher.1989}. To capture this effect and analyze how infiltrated oxygen influences surface reaction pathways, we replace a fraction of the bulk nitrogen atoms with oxygen to achieve a final bulk Si:N:O atomic ratio of 9:10:1. Accordingly, the final bulk compositions are Si$_{54}$O$_{108}$ for a-SiO$_2$ and Si$_{72}$O$_{8}$N$_{80}$ for a-SiN$_x$.

All amorphous surfaces are constructed using a melt-quench simulation protocol. Initially, crystalline supercells, specifically $3\times3\times2$ for $\alpha$-SiO$_2$ and $2\times2\times3$ for $\beta$-Si$_3$N$_4$, are prepared to match the aforementioned experimental densities and stoichiometries. To eliminate crystalline ordering, each supercell undergoes a pre-melting stage at 6000 K for $\alpha$-SiO$_2$ and 5000 K for $\beta$-Si$_3$N$_4$. Subsequently, structures are melted at temperatures of 3000~K for 10~ps ($\alpha$-SiO$_2$) and 4000~K for 10~ps ($\beta$-Si$_3$N$_4$). The molten systems are then quenched from these melting temperatures to 0~K over 15~ps. After quenching, structural relaxation yields stable amorphous bulk configurations. Following this procedure, we obtain 3 distinct bulk structures, each for a-SiO$_2$ and a-SiN$_x$ with a small number of oxygen defects. In the case of a-SiN$_x$, N$_2$ gas occasionally evolves during high-temperature melting, as reported in a previous computational study~\cite{Kang.2018}. We remove these evolved gas molecules before the relaxation step.

From these bulk structures, amorphous surfaces are created by cleaving each bulk structure along the $z$-axis, followed by introducing a vacuum region of 15~\AA\ to avoid spurious interactions between periodic images. For a-SiO$_2$, cleavage planes are selected randomly, while for a-SiN$_x$, cleavage is performed near oxygen atoms to analyze the reactivity of surface oxygen groups induced by oxygen impurities.
Each bulk structure yields two distinct surface terminations, resulting in a total of six unique surfaces for each material. Dangling bonds on the cleaved surfaces are then passivated using functional groups (\ce{-OH}, \ce{-H}, \ce{-NH2}, and \ce{=NH}) to closely replicate experimentally determined functional-group densities (see Table~\ref{tab:passivation}), thereby restoring the original bulk coordination numbers of the surface atoms. We intentionally include $\ce{=NH}$ and $\ce{-H}$ termini in the initial passivation, even though these groups are uncommon on fully relaxed amorphous surfaces. Their inclusion serves two purposes: promoting bridge formation during subsequent annealing and maintaining surface stoichiometry close to 1.33. While previous computational studies of amorphous \ce{SiO2} surfaces have employed similar melt-quench protocols~\cite{Ewing.2014,Tielens.2008,Ugliengo.2008,Khumaini.2024,Siron.2023,Caro.2018,Sandupatla.2015}, our passivation strategy differs in that we incorporate these additional functional groups (\ce{=NH} and \ce{-H}) and employ a subsequent annealing procedure to promote chemically realistic surface rearrangements. A schematic illustration of the complete modeling protocol is provided in Figure S1.

Following the passivation, the slabs undergo annealing at 1000~K for 5~ps, then quenching to 0~K over an additional 5~ps. During this annealing process, initial \ce{Si=NH} groups rearrange into bridging \ce{Si-NH-Si} structures within the amorphous silicon nitride network. Simultaneously, H atoms released from \ce{Si-H} bonds passivate nearby dangling bonds, forming \ce{-OH}, \ce{-NH2}, and similar functional groups. A final structural relaxation step yields stable amorphous surface configurations. Analyses of surface reactive sites carried out for the resulting surfaces are summarized in Figure~\ref{Figures/Sites.pdf} (the surface speciation analysis algorithm provided in supplementary information S3). As illustrated in Figure~\ref{Figures/Sites.pdf}b, the a-SiO$_2$ surface primarily features two reactive site types: silanol (\ce{-OH}) and siloxane (\ce{-O-}). In contrast, the a-SiN$_x$ surface predominantly contains amine (\ce{-NH2}) and bridging imide (\ce{-NH-}) sites, along with oxygen-containing reactive sites (\ce{-OH} and \ce{-O-}), as shown in Figure~\ref{Figures/Sites.pdf}c.

For the generated a-SiO$_2$ slabs, the surface density of silanol groups is 6.1~nm$^{-2}$, slightly higher than typical experimental values around 4.5~nm$^{-2}$~\cite{Zhuravlev.2000,Dugas.2003}. This difference can be attributed to distinct methodologies: the referenced experimental techniques mainly quantify readily reactive $\ce{-OH}$ sites by adsorbing specific molecules, whereas our computational methods account for all silanol groups exposed on the topmost surface, including both those fully exposed on the outermost plane and subsurface groups that exhibit limited reactivity. On the other hand, analysis of the hydrogen-bonding configuration shows a density of 3.9 nm$^{-2}$ for vicinal and 2.2 nm$^{-2}$ for isolated silanol groups. This indicates that the density of vicinal silanol groups is nearly 1.5 times that of isolated groups, consistent with both experimental findings~\cite{Zhuravlev.2000,Dugas.2003} and computational predictions~\cite{Ewing.2014} (the algorithm for classifying vicinal and isolated groups are provided in supplementary information S3). While we identify the presence of \ce{Si-H} groups on the surface, their experimental detection is rarely reported. These groups exhibit negligible densities (typically below 1 nm$^{-2}$) in our calculations. Therefore, we neglect the reactivity of these groups. For the a-SiN$_x$ slabs, to the best of our knowledge, experimental data on specific surface densities or hydrogen-bonding configurations of \ce{-NH-} or \ce{-NH2} groups are currently unavailable. In this study, we employ the passivation scheme to generate diverse surface groups, resulting in approximate surface densities of 4 nm$^{-2}$ for both \ce{-NH2} and \ce{-NH-}. Should future experimental characterizations provide quantitative surface-group densities and hydrogen-bonding configurations for PECVD-grown a-SiN$_x$ surfaces, integrating these data into our modeling would enable more realistic simulations.

\begin{table}[htbp]
    \centering
    \caption{Passivation scheme for surface dangling bonds, classified by element type and dangling-bond count after bulk cleavage.}
    \label{tab:passivation}
    \begin{tabular}{cccc}
        \toprule
        Surface & Element & \makecell{Number of\\dangling bonds} & Passivation \\
        \midrule
        \multirow{4}{*}{SiO$_2$} 
            & Si & 3 & Si $\rightarrow$ Si(OH)$_2$H \\
            & Si & 2 & Si $\rightarrow$ Si(OH)H \\
            & Si & 1 & Si $\rightarrow$ SiOH \\
            & O  & 1 & O $\rightarrow$ OH \\
        \midrule
        \multirow{6}{*}{SiN$_x$}
            & Si & 3 & Si $\rightarrow$ Si(NH)H \\
            & Si & 2 & Si $\rightarrow$ SiNH \\
            & Si & 1 & Si $\rightarrow$ SiNH$_2$ \\
            & N  & 2 & N $\rightarrow$ NH$_2$ \\
            & N  & 1 & N $\rightarrow$ NH \\
            & O  & 1 & O $\rightarrow$ OH \\
        \bottomrule
    \end{tabular}
\end{table}

To compare the reactivity of amorphous and crystalline surfaces, we also model crystalline SiO$_2$ (c-SiO$_2$) and crystalline Si$_3$N$_4$ (c-Si$_3$N$_4$) surfaces, which are fully passivated with OH and NH$_2$ groups, respectively. For c-SiO$_2$, we employ an $\alpha$-quartz structure, and a 4$\times$4$\times$2 supercell is cleaved along the $z$-axis. For c-Si$_3$N$_4$, we use a $\beta$-Si$_3$N$_4$ structure, and a 2$\times$2$\times$3 supercell is cleaved along the same orientation. The resulting surfaces are appropriately passivated to yield stable, fully passivated configurations.

In crystalline structures, bridge sites are primarily located within subsurface regions, requiring substantial energy to break the crystalline order during bridge cleavage reactions. Although it is possible to construct surface-exposed bridge-site models for crystalline surfaces, such models introduce the formation of unrealistic \ce{Si-H} groups or strained bonding configurations~\cite{Ande.2015}. To compare the reactivity of crystalline and amorphous surfaces without such artifacts, we model only OH- and NH$_2$-terminated crystalline surfaces.

\fullsizefigure{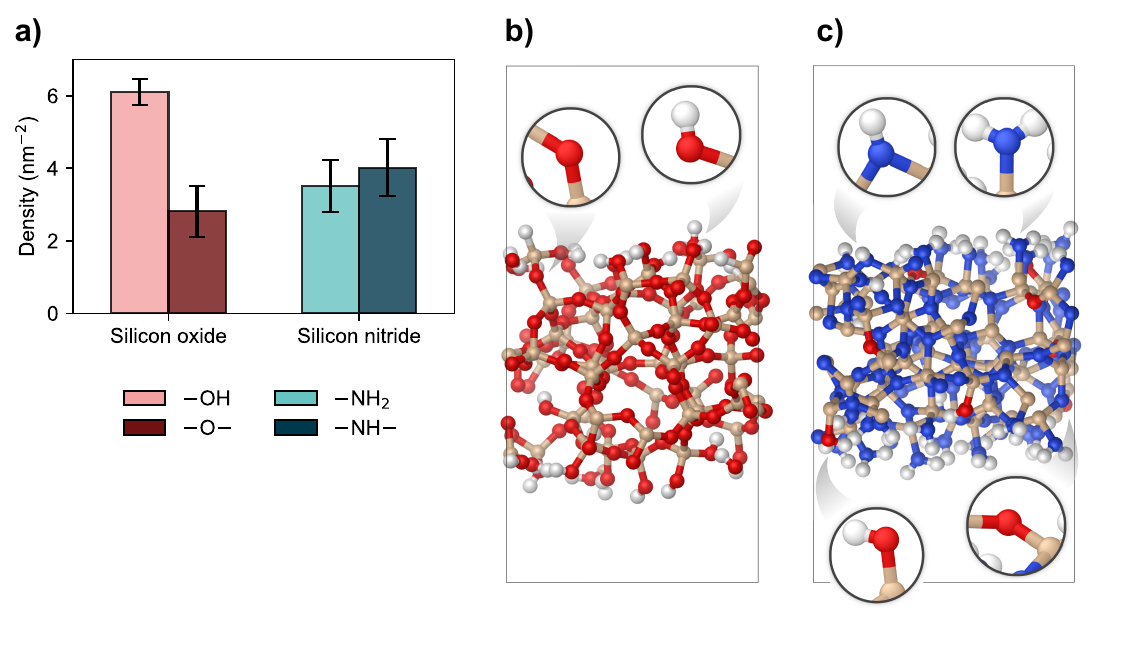}{(a) Average surface reactive-site densities for amorphous \ce{SiO2} (a-\ce{SiO2}) and amorphous \ce{SiN}$_x$ (a-\ce{SiN}$_x$) surfaces, calculated from six surfaces for each material. Error bars indicate standard deviations. Example atomic structures of (b) a-\ce{SiO2} and (c) a-\ce{SiN}$_x$, with insets highlighting major reactive sites on each surface.} 

\subsection{Reactivity of inhibitors on the surfaces}

We investigate the chemical reactivity of surface functional groups toward DMATMS and ETS by computing reaction energies ($\Delta E_\mathrm{r}$) and activation energies ($E_\mathrm{a}$), defined as follows:

\begin{equation}
\Delta E_{\mathrm{r}} = E_{\mathrm{chemisorption}} - E_{\mathrm{physisorption}}
\end{equation}
\begin{equation}
E_\mathrm{a} = E^\ddagger - E_{\mathrm{physisorption}}
\end{equation}
where $E_{\mathrm{chemisorption}}$ is the energy of the chemisorbed state, $E_{\mathrm{physisorption}}$ is the energy of the physisorbed state, and $ E^\ddagger$ is the energy of the transition state. In the following subsections, we propose plausible reaction pathways for DMATMS and ETS and evaluate their reactivity based on calculated $\Delta E_\mathrm{r}$ and $E_\mathrm{a}$. We specifically consider reactions at the \ce{-OH} and \ce{-O-} sites on \ce{SiO2} surfaces, as well as the \ce{-NH2} and \ce{-NH-} sites on \ce{SiN}$_x$ surfaces. 

Reactions of an inhibitor molecule on the surface follow the typical ALD precursor reaction mechanism due to the structural similarity of their reactive head groups. The reaction initiates with physisorption, where surface hydrogen atoms form hydrogen bonds with the inhibitor molecule. Subsequently, proton transfer occurs from the surface to the inhibitor, accompanied by the attachment of the cleaved inhibitor fragment to the surface and the desorption of the remaining molecular fragment as a gaseous byproduct. This proton-mediated ligand exchange mechanism, commonly observed in ALD precursor chemistry~\cite{Richey.2020}, results in chemisorption of the inhibitor onto the surface. Our proposed reaction pathways consistently prioritize such proton transfer reactions, as the Brønsted acid-base nature of surface reactive sites and inhibitors makes this type of reaction highly favorable.

At the terminal sites of $\ce{-OH}$ and $\ce{-NH2}$, proton transfer can occur bidirectionally, either from the surface groups to the inhibitor molecules or vice versa. The specific location where the transferred proton attaches determines the resulting gaseous byproducts, such as $\ce{H2O}$, $\ce{NH3}$, or $\ce{HNR2}$. For bridging $\ce{-O-}$ and $\ce{-NH-}$ sites, proton transfer is also the primary reaction principle. However, with the exception of the $\ce{-NH-}$ group providing atoms to the inhibitor, these reactions generally result only in bond cleavage without the formation of gaseous byproducts. In cases where multiple reaction pathways are possible at the same surface site, such as $\ce{-NH-}$, we have prioritized the most thermodynamically stable (exothermic) pathway. The reactivities of these alternative reaction pathways are discussed in supplementary information S5.

We systematically compare reaction mechanisms and corresponding reactivities on both amorphous and crystalline surfaces. For amorphous surfaces, we select at least three distinct sites per functional group type with minimal steric interference, testing multiple physisorption configurations at each site. The resulting energies from these representative sites are then averaged for presentation. By contrast, for crystalline surfaces, calculations are performed at a single representative reactive site owing to their structural symmetry. Detailed sampling methodology and considerations for hydrogen-bond alignment are provided in supplementary information S4.

\subsubsection{Reaction on silicon oxide}\label{reaction_sio2}

Figure~\ref{Figures/Reaction_SiO2.pdf} presents favorable reaction pathways of DMATMS and ETS on \ce{SiO2} surfaces, with their corresponding reactivities summarized in Figure~\ref{Figures/SiO_E.pdf}. At \ce{-OH} sites, both inhibitors follow a similar reaction mechanism. The surface \ce{-OH} site releases its proton, leading to the formation of surface-bound \ce{SiO^*-SiR3} species (where * denotes a surface site; Figures~\ref{Figures/Reaction_SiO2.pdf}a and c). During this process, the released proton transfers to a basic site on the precursor fragment. This results in the evolution of gaseous byproducts: \ce{HNR2} for DMATMS and \ce{H2O} for ETS. Both DMATMS and ETS exhibit exothermic reactions with \ce{-OH} sites on a-\ce{SiO2} (Figure~\ref{Figures/SiO_E.pdf}). In contrast, the resulting $E_\mathrm{a}$ indicates comparatively lower reactivity on crystalline surfaces, where $E_\mathrm{a}$ is higher by 0.1--0.2 eV for both molecules. This highlights an increased reactivity on amorphous surfaces compared to their crystalline surfaces. On the other hand, DMATMS demonstrates a lower $E_\mathrm{a}$ of 0.48 $\pm$ 0.16 eV compared to ETS, whose $E_\mathrm{a}$ exceeds 1 eV. The differences in reactivity between DMATMS and ETS at \ce{-OH} sites originate primarily from the differences in bond strength of their respective head groups. The weaker Si-N bond in DMATMS (compared to the Si-O bond in ETS) facilitates bond breaking, resulting in lower $E_\mathrm{a}$. These computational results agree well with previous theoretical analysis of c-\ce{SiO2} surfaces, which similarly identified lower $\Delta E_\mathrm{r}$ and $E_\mathrm{a}$ values for aminosilane interactions with \ce{-OH} sites~\cite{Lee.2021s4}.

\halfsizefigure{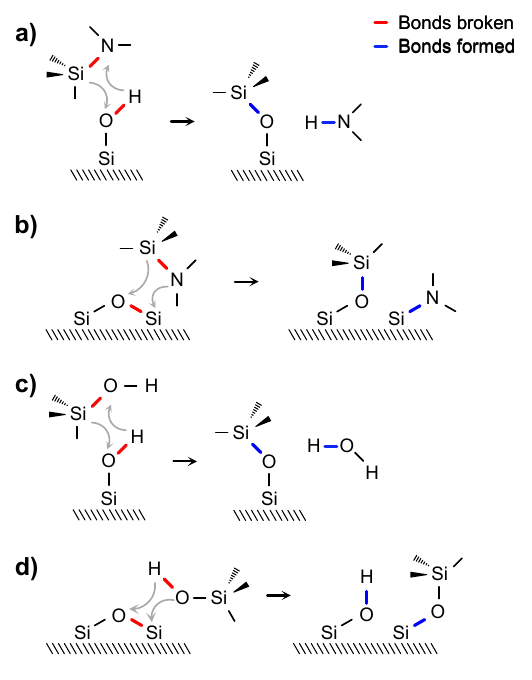}{Reaction pathways of DMATMS and ETS at reaction sites \ce{Si-OH} and \ce{Si-O-Si} on \ce{SiO2}. (a) Reaction of DMATMS at a \ce{Si-OH} site. (b) Reaction of DMATMS at a \ce{Si-O-Si} site. (c) Reaction of ETS at a \ce{Si-OH} site. (d) Reaction of ETS at a \ce{Si-O-Si} site.} 

\halfsizefigure{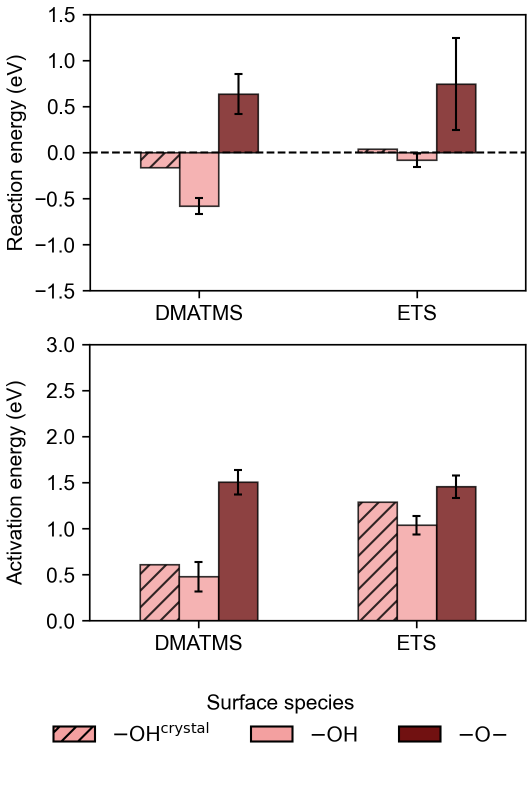}{Reaction energies ($\Delta E_\mathrm{r}$, upper) and activation energies ($E_\mathrm{a}$, lower) regarding the reaction sites on \ce{SiO2} surface. Hatch patterns indicate reactions on crystalline \ce{SiO2} surfaces.}

In siloxane bridge (\ce{-O-}), the reaction pathways involve cleavage of a \ce{Si-O} bond, as illustrated in Figures~\ref{Figures/Reaction_SiO2.pdf}b and d. During this process, the cleaved surface Si atom bonds to the inhibitor nucleophilic sites (N or O), forming \ce{Si^*-NR2} for DMATMS and \ce{Si^*-OH} for ETS. Simultaneously, the cleaved O atom forms a new bond with the inhibitor Si, creating a \ce{SiO^*-SiR3} or \ce{Si^*-OSiR(OH)2}. This process induces significant structural distortion in the surface due to both the cleavage of surface bridge bonds and the steric hindrance caused by the additional chemisorption of inhibitor fragments. As a result, as shown in Figure~\ref{Figures/SiO_E.pdf}, reactions at siloxane sites are thermodynamically unfavorable (endothermic) and exhibit a high $E_\mathrm{a}$ of $\sim$1.5 eV. These unfavorable energetics lead to relatively low reactivity for both DMATMS and ETS on siloxane sites.

\subsubsection{Reaction on silicon nitride}

Similar to the previous section \ref{reaction_sio2}, Figure~\ref{Figures/Reaction_Si3N4.pdf} shows favorable reaction pathways of DMATMS and ETS on SiN$_x$ surface. In addition, Figure~\ref{Figures/SiON_E.pdf} illustrates the calculated reactivity for the possible reaction pathways of DMATMS and ETS on \ce{SiN}$_x$. The reactivity at oxygen species (\ce{-OH} and \ce{-O-}) on the a-SiN$_x$ shows values comparable to those on the a-\ce{SiO2}. Detailed analysis of these reaction pathways is discussed in supplementary information S6. 

\halfsizefigure{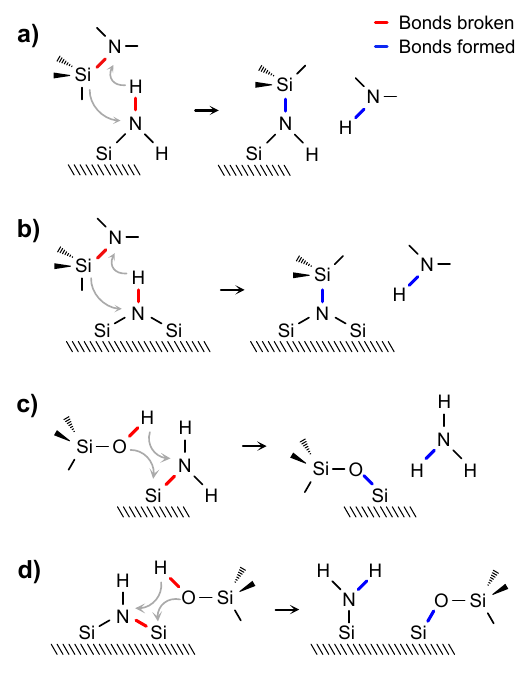}{Reaction pathways of DMATMS and ETS at reaction sites \ce{-NH2} and \ce{-NH-} on \ce{SiN}$_x$. (a) Reaction of DMATMS at a $\ce{-NH2}$ site. (b) Reaction of DMATMS at a $\ce{-NH-}$ site. (c) Reaction of ETS at a $\ce{-NH2}$ site. (d) Reaction of ETS at a $\ce{-NH-}$ site.}

\halfsizefigure{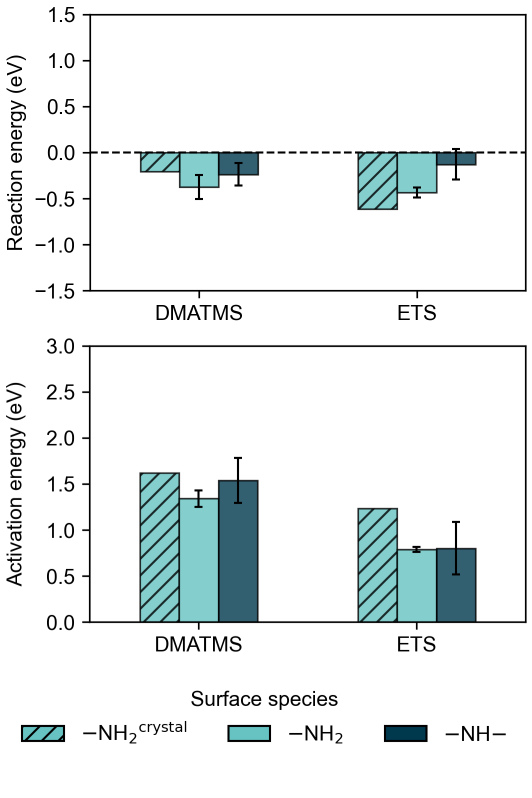}{Reaction energies ($\Delta E_\mathrm{r}$, upper) and activation energies ($E_\mathrm{a}$, lower) regarding the reaction sites on \ce{SiN}$_x$ surface. Hatch patterns indicate reactions on crystalline \ce{SiN}$_x$ surfaces.}

The reaction of DMATMS at \ce{-NH2} surface sites forms surface-bound \ce{SiN^*H-SiR3} species and releases gaseous \ce{HNR2} as a byproduct (Figure~\ref{Figures/Reaction_Si3N4.pdf}a). Our calculations reveal that DMATMS exhibits considerably higher reactivity at amorphous \ce{-NH2} sites ($E_\mathrm{a}$ = 1.34 $\pm$ 0.09 eV) compared to crystalline sites ($E_\mathrm{a}$ = 1.62 eV), as shown in Figure~\ref{Figures/SiON_E.pdf}. This difference highlights the importance of employing amorphous surface models for accurately predicting molecular selectivity. While previous DFT studies reported lower $E_\mathrm{a}$ ($\sim 1.0$ eV) for other aminosilane precursors on c-\ce{SiN}$_x$ surfaces~\cite{Murray.2014,Huang.2014}, the higher $E_\mathrm{a}$ calculated for DMATMS can be attributed to increased steric hindrance from its bulkier tail group.

In addition, ETS exhibits higher reactivity at \ce{-NH2} sites on amorphous surfaces with $E_\mathrm{a}$ of 0.79 $\pm$ 0.03 eV and 1.23 eV on crystalline surfaces (Figure~\ref{Figures/SiON_E.pdf}). In these cases, the reaction primarily produces a \ce{Si^*-OSiR(OH)2} surface bond and \ce{NH3} as a byproduct (Figure~\ref{Figures/Reaction_Si3N4.pdf}c). Although ETS can alternatively yield \ce{SiN^*H-SiR(OH)2} with the evolution of \ce{H2O} (see Figure S2), the \ce{NH3}-producing pathway is thermodynamically favored due to its exothermic nature on both amorphous and crystalline surfaces. Consequently, the energetically favorable, exothermic reaction producing \ce{NH3} is presented in Figure~\ref{Figures/SiON_E.pdf}. From an acid-base chemistry perspective, ETS, being more acidic, is predicted to exhibit higher reactivity with basic \ce{-NH2} surface sites compared to DMATMS, consistent with our computational findings.

At the \ce{-NH-} site, DMATMS forms a bridged surface configuration, denoted as \ce{Si-(N^*-SiR3)-Si}, without breaking the existing nitrogen bridge (Figure~\ref{Figures/Reaction_Si3N4.pdf}b). This reaction occurs through a proton trasfer, where the hydrogen on the bridging nitrogen is replaced by the inhibitor fragment (\ce{SiR3}). Since reactions of DMATMS at both \ce{-NH2} and \ce{-NH-} involve similar bond-forming (\ce{Si-N} and \ce{N-H}) and bond-breaking steps (\ce{N-H}), the $E_\mathrm{a}$ for DMATMS at these two nitrogen-containing sites are comparable (1.34 $\pm$ 0.09 eV and 1.54 $\pm$ 0.25 eV, respectively). ETS reaction at \ce{-NH-} sites leads to cleavage of a bridge bond, producing \ce{SiN^*-H2} and \ce{Si^*-OSiR}. Although an alternative reaction pathway producing \ce{H2O} and \ce{Si-(N^*-SiR(OH)2)-Si} without cleaving the bridge bond is possible, it exhibits endothermic reaction energy and $\sim$0.5 eV higher $E_\mathrm{a}$ (see Figure S3). Therefore, we represent the more favorable bridge-cleaving reaction pathway in Figure~\ref{Figures/SiON_E.pdf}, which is exothermic and has $E_\mathrm{a}$ of 0.80 $\pm$ 0.28  eV for the \ce{-NH-} site. 

In summary, we compare multiple reaction pathways of DMATMS and ETS across various sites on a-SiO$_2$ and a-SiN$_x$, with particular emphasis on higher reactivity of amorphous surfaces relative to crystalline surfaces. The results underscore the importance of employing amorphous surface modeling to capture realistic surface-inhibitor interactions, providing a deeper understanding of inhibitor selectivity in materials science and surface engineering contexts.

\section{Discussion}
We have demonstrated that different amorphous sites exhibit varying reactivity toward inhibitor molecules. In particular, we evaluated the reactivity at both terminal sites (\ce{-OH} and \ce{-NH2}) and bridge sites (\ce{-O-} and \ce{-NH-}). For highly selective AS-ALD, the NGS must ideally have all its reactive sites passivated. For example, highly reactive molecules such as DMAI or TMA can adsorb effectively at bridge sites on a-\ce{SiO2} surfaces~\cite{Wellmann.2024, Xu.202223f, Sandupatla.2015}. However, the inhibitors studied here, DMATMS and ETS, are not ideal in this context, as both exhibit endothermic reaction pathways at \ce{-O-} and \ce{-NH-} sites and therefore fail to passivate the bridge sites on a-\ce{SiO2} and a-\ce{SiN}$_x$. Nevertheless, complete passivation of bridge sites is not always necessary---particularly when the target precursors are less reactive than inhibitors. For instance, diisopropylaminosilane (DIPAS), a common precursor structurally analogous to the aminosilane DMATMS, exhibits limited reactivity toward \ce{-O-} bridge sites. In this scenario, DMATMS may serve effectively as a selective inhibitor, preferentially targeting \ce{Si-OH} sites on a-\ce{SiO2} during ALD processes involving DIPAS-like precursors.

Considering these scenarios, we propose the following screening procedure for identifying effective inhibitors on binary surfaces:

\begin{enumerate}
\item Identify potential surface reactive sites on both substrates and evaluate their reactivity with the target precursor.
\item Select inhibitor molecules that adsorb effectively on the NGS sites exhibiting high precursor reactivity.
\item Ensure the inhibitor exhibits low reactivity with sites on the GS that are reactive to the precursor.
\end{enumerate}
This approach can be extended to large-scale screening of precursor–inhibitor combinations across various surface systems. 

\section{Conclusion}
We investigated the reaction pathways of DMATMS and ETS, evaluating their reactivities on amorphous and crystalline surfaces of silicon oxide and silicon nitride. DMATMS exhibited higher reactivity on silicon oxide surfaces, while ETS showed preferential reactivity toward silicon nitride surfaces. Both inhibitors demonstrated enhanced reactivity at terminal surface sites (\ce{-OH} and \ce{-NH2}) on amorphous surfaces compared to their crystalline counterparts. For bridge sites on oxides and nitrides, multiple reaction pathways are possible, with the cleavage of the bridge being the most probable pathway, except in the case of DMATMS reacting with nitride. The reactivity of DMATMS with \ce{-NH} was found to be comparable to that with \ce{-NH2}, with both reactions producing volatile products. These findings underscore the critical role of amorphous surface modeling in accurately evaluating the reactivity and adsorption behavior of inhibitor molecules on practical target surfaces. Furthermore, we describe a computational screening procedure that accounts for site-specific reactivity toward precursors and inhibitors, facilitating efficient and rational theoretical design of AS-ALD precursor–inhibitor combinations.

\section{Author contributions: CRediT}
    \textbf{Gijin Kim:} Conceptualization, Data Curation, Formal Analysis, Investigation, Methodology, Software, Validation, Visualization, Writing – original draft. 
    \textbf{Purun-hanul Kim:} Conceptualization, Methodology, Validation, Writing – review \& editing. 
    \textbf{Suk Gyu Hahm:} Conceptualization. 
    \textbf{Myongjong Kwon:} Conceptualization. 
    \textbf{Byungha Park:} Conceptualization. 
    \textbf{Changho Hong:} Conceptualization, Methodology, Validation, Writing – review \& editing. 
    \textbf{Seungwu Han:} Conceptualization, Project Administration, Resources, Supervision.

\section{Acknowledgements}
    This work was supported by the National Research Foundation of Korea (NRF) grant funded by the Korea government (MSIT) (RS-2025-25442273) and Samsung Electronics under grant code IO221104-03413-01. Computational resources were provided by Center for Advanced Computations (CAC) at Korea Institute for Advanced Study (KIAS) and Korea Institute of Science and Technology Information (KISTI) National Supercomputing Center (KSC-2023-CRE-0337).

\appendix

\section{Supplementary material}
\label{sec:sample:appendix}
Supplementary data to this article can be found online at ...

 \bibliographystyle{elsarticle-num} 
 \bibliography{Basic,2211_AS-ALD}





\end{document}